\documentclass[3p, 11pt]{elsarticle} 

\usepackage{lineno}

\usepackage{graphicx,subfigure}
\usepackage{lineno,hyperref}
\usepackage{tabularx}
\usepackage{xspace}
\usepackage{booktabs}

\usepackage[section]{placeins}

\makeatletter
\def\ps@pprintTitle{%
 \let\@oddhead\@empty
 \let\@evenhead\@empty
 \def\@oddfoot{}%
 \let\@evenfoot\@oddfoot}
\makeatother

\begin{document}

\begin{frontmatter}

\title{Dose rate dependence of TID damage to 65 nm CMOS transistors in X-ray irradiations of the ATLAS ITk Pixel ASIC (ITkPix)}


\author[a]{D. Bortoletto}
\author[b]{A. Dimitrievska}
\author[c]{M. Garcia-Sciveres}
\author[c]{T. Heim}
\author[c]{M. Mironova}
\author[a]{R. Plackett}
\author[a]{I. Shipsey}
\author[c]{J. Xiong}

\address[a]{University of Oxford (UK)}
\address[b]{University of Birmingham (UK)}
\address[c]{LBNL (US)}

\begin{abstract}

The ATLAS Inner Tracker (ITk) upgrade for the High-Luminosity LHC (HL-LHC) requires a radiation-tolerant pixel readout chip, which must withstand a total ionising dose (TID) of up to 1 Grad. The readout ASIC for the ITk upgrade has been designed by the RD53 collaboration using 65 nm CMOS technology. In order to characterise the radiation tolerance of the chip digital logic, the RD53 ASICs include ring oscillators, which can be used to measure gate delay degradation. Extensive X-ray irradiation studies of the ring oscillators have been performed on the ITk Pixel pre-production readout ASIC, ITkPixV1. A dependence of radiation damage on dose rate has been observed in 65 nm CMOS technology. This paper aims to quantify the dose rate dependence of TID damage to the ITkPix ring oscillators and, therefore, the ITkPix ASIC digital logic. X-ray irradiations at different dose rates between 20 krad/h and 30 Mrad/h are compared. A dose rate dependence is observed, with 2-3 times more damage at the lowest dose rate of 20 krad/h, compared to 4 Mrad/h. The dose rate dependence was also observed to be dependent on transistor size and type.

\end{abstract}

\end{frontmatter}


\section{Introduction}

The ATLAS experiment will upgrade its inner detector for operation during High-Luminosity LHC (HL-LHC) to cope with the increased instantaneous and total luminosity and increased collisions per beam-crossing. New pixel readout chips were designed by the RD53 collaboration for the ATLAS and CMS experiments to satisfy the harsher requirements on the radiation tolerance and the efficient and fast detection of signals. Following the RD53 design effort, the ITk pre-production ASIC ITkPixV1 was submitted in 2020, followed by the final production ASIC ITkPixV2 in 2023. 

The readout chip is required to withstand a total ionising dose (TID) of up to 1 Grad in the innermost ITk layers, and the radiation tolerance of the chip has to be validated experimentally \cite{RadDamage}. For radiation testing, the chip includes 42 ring oscillators, using different types and sizes of logic gates, which allow the characterisation of gate delay increase with irradiation. Generally, TID damage can occur in different parts of transistors, and for the small transistors in 65 nm technology, channel edge effects make up the most significant contribution \cite{EdgeEffects}. For instance, charge build-up in the Shallow Trench Isolation (STI) and spacers can significantly change the leakage current and threshold voltage of the transistor, phenomena known as Radiation-Induced Short Channel (RISCE) and Narrow Channel (RINCE) effects \cite{RINCE,Spacers}. Irradiation studies of single transistors show that the radiation tolerance is worse for narrow-channel transistors, and more damage is observed during irradiations using low dose rates and at higher temperatures \cite{DoseRate, LDR}.

This dose rate dependence needs to be considered when performing radiation studies for ITk. While it is possible to achieve the expected ITk TID of 1 Grad on the timescale of weeks when irradiating at dose rates of several Mrad/h, the dose during ITk operation will be delivered at a much lower dose rate. The highest  dose rate in the innermost ITk layers is expected to be around 20 krad/h. Therefore, understanding the scaling of TID damage with dose rate is crucial to determine whether the ITkPix ASIC is sufficiently radiation tolerant to withstand the ITk environment. This paper aims to quantify the dose rate dependence of TID damage in small transistor gates by performing X-ray irradiation campaigns of the ITkPixV1 ring oscillators at different doses ranging from 20 krad/h to 30 Mrad/h and comparing the gate delay degradation at the different dose rates, after total doses of at least 10 Mrad. 

All measurements presented in this paper have been performed using the pre-production readout chip ITkPixV1. 

\section{Setup}

This section introduces the ring oscillators in the ITkPixV1 chip, which are used to perform the radiation studies, along with the X-ray irradiation setup and calibration measurements. 

\subsection{Ring oscillators in the ITkPixV1 chip}

The ITkPixV1 chip includes various ring oscillators, which can be used to characterise radiation damage to different kinds of gates used in the chip digital logic. These ring oscillators are made with varying types of logic cells and using a range of transistor sizes. Each oscillator drives a 12-bit counter, while a 4-bit counter counts the number of start/stop pulses received. From this, it is possible to calculate the frequency of the ring oscillator and the gate delay.

Table \ref{tab:ROSC_labels} summarises the different kinds of ring oscillators in ITkPixV1, their transistor drive strength and the length of the oscillators (i.e. the number of gates in each circuit). The transistor drive strength is related to the size of some of the transistors in the circuit, scaling as $W/L$ for those transistors, where $W$ and $L$ are the transistor widths and lengths, respectively. In particular, strength 0 gates use minimum size transistors, while strength $>$0 gates do not have minimum size transistors.


\begin{table}
  \centering
  \caption{Summary of the 42 ring oscillators available in ITkPixV1 with their gate type, drive strength and length.}
  \small
  \begin{tabular}{clll}
    \toprule
    Ring oscillator number & Type & Strength & Length \\
    \midrule
    \midrule
    \multicolumn{4}{c}{Bank A} \\
    \midrule
    0 & Inv. CLK driver & 0 & 55 \\
    1 & Inv. CLK driver & 4 & 51 \\
    2 & Inverter & 0 & 55 \\
    3 & Inverter & 4 & 55 \\
    4 & 4-input NAND & 0 & 19 \\
    5 & 4-input NAND & 4 & 19 \\
    6 & 4-input NOR & 0 & 19 \\
    7 & 4-input NOR & 4 & 19 \\
    \midrule
    \multicolumn{4}{c}{Bank B left/right} \\
    \midrule
    0 \& 1 & Inv. CLK driver & 0 & 38.2 \\
    2 \& 3 & Inv. CLK driver & 4 & 44.5 \\
    4 \& 5 & Inverter & 0 & 38.1 \\
    6 \& 7 & Inverter & 4 & 44.3 \\
    8 \& 9 & 4-input NAND & 0 & 12.6\\
    10 \& 11 & 4-input NAND & 4 & 16 \\
    12 \& 13 & 4-input NOR & 0 & 14.5 \\
    14 \& 15 & 4-input NOR & 4 & 14.5 \\
    \midrule
    \multicolumn{4}{c}{Bank B FF} \\
    \midrule
    16 \& 17 & Scan D-flip-flop & 0 & 6.1 \\
    18 \& 19 & D-flip-flop & 1 & 6.2 \\
    20 \& 21 & Neg. edge D-flip-flop & 1 & 5 \\
    \midrule
    \multicolumn{4}{c}{Bank B LVT} \\
    \midrule
    22 & LVT inverter & 0 & 40.6\\
    23 & LVT inverter & 4 & 56 \\
    24 & LVT 4-input NAND & 0 & 16.5 \\
    25 & LVT 4-input NAND & 4 & 22.8 \\
    \midrule
    \multicolumn{4}{c}{Bank B CAPA} \\
    \midrule
    26-33 & Inj-cap loaded 4-input NAND & 4 & 16.8 \\
    \bottomrule

  \end{tabular}
  \label{tab:ROSC_labels}
\end{table}

For each ring oscillator, it is possible to count the number of oscillations in a given time, which then can be used to determine the ring oscillator frequency $f$, or the delay of the individual gates as $T=1/(N\times f)$, where $N$ is the length of the ring oscillator. It is also possible to calculate the relative increase in gate delay as $\Delta T=1/(N\times f) - 1/(N\times f_{\mathrm{start}})$, where $f_{\mathrm{start}}$ is the ring oscillator frequency before irradiation. The gate delay will increase with increasing radiation damage, and the ring oscillator frequency will decrease, which is the primary measure of radiation damage to the digital logic. Simulations of the chip determined that the gate delay increase has to stay below 200\% to ensure sufficiently good timing in the digital logic of the chip. Before irradiation, the logic gate delays are between 0.03 to 0.4 ns.

\subsection{X-ray irradiation setup and calibration}

Most irradiation campaigns presented in this paper were performed at the Oxford Microstructure Detector facility (OPMD) using a Euroteck X-ray system \cite{euroteck_2020}. The X-rays are produced with a tungsten target and then filtered by a 0.8 mm Beryllium window when exiting the tube. The tube can be operated with an anode voltage of 5 kV to 60 kV and a current between 1 and 60 mA, with the maximum achievable power being 3 kW. For the ITkPixV1 irradiations, the X-rays are additionally filtered by 150 $\mu m$ of Aluminium to remove low-energy components. The dose rate (in $SiO_2$) delivered by the X-ray tube at a particular setting was measured before each irradiation campaign using a diode calibrated at CERN \cite{calib}. 

\begin{figure}
\centering
  \centering
  \includegraphics[width=0.45\textwidth]{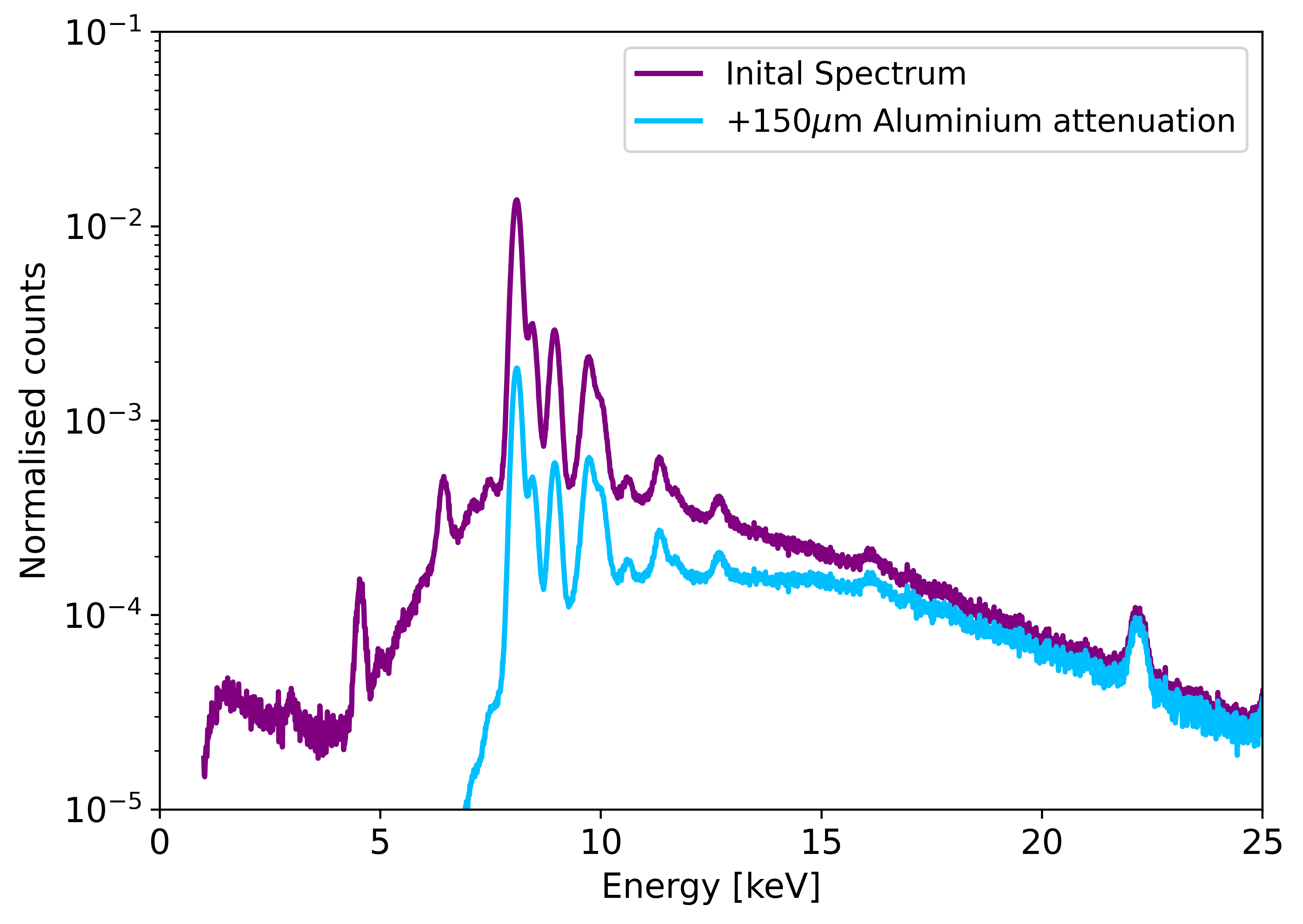}
  \includegraphics[width=0.45\textwidth]{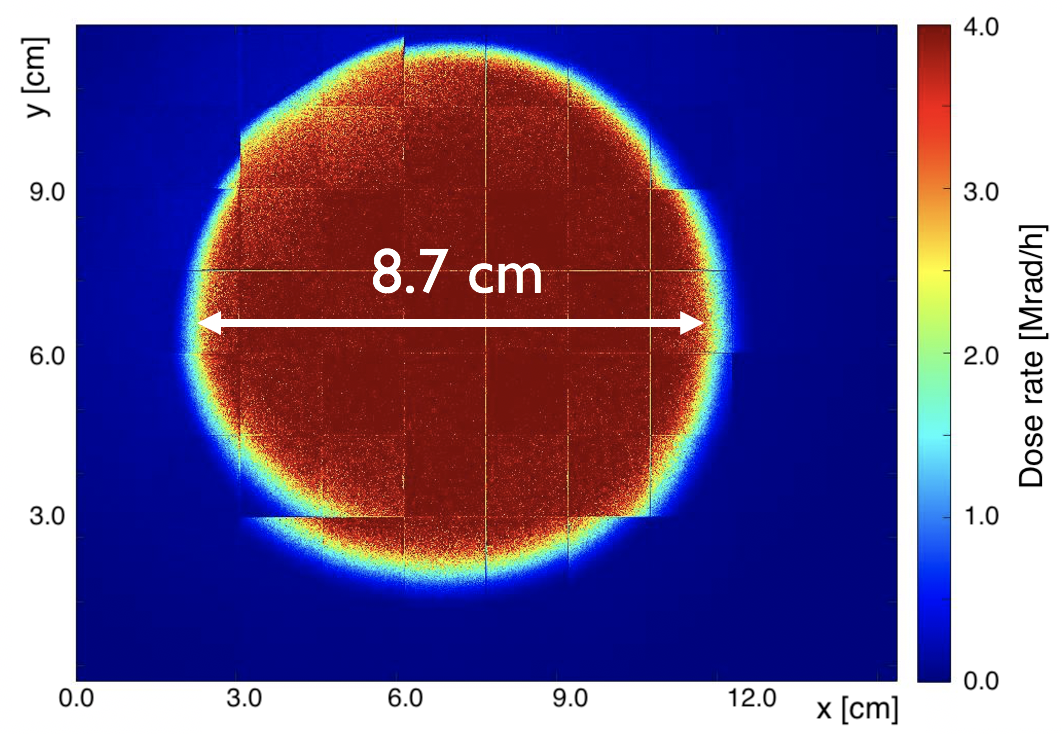}\qquad
  \caption{Energy spectrum of the X-ray tube at an operational voltage of 40 kV (left), with and without a $\mathrm{150\;\mu m}$ Aluminium filter. Beam spot 10 cm from the exit of the tube (right). }
  \label{fig:xray1}
\end{figure}

The X-ray tube was further characterised by measuring its spectrum with a silicon drift detector, with the results shown in Figure  \ref{fig:xray1}, with and without the Aluminium filter. The beam spot profile was measured using a Medipix Merlin detector \cite{Merlin}, by counting the number of photon hits per $\mathrm{10\;\mu s}$ window at a distance of 10 cm from the tube exit. The resulting map of hits was then scaled to the measured dose rate at this distance from the tube and is also shown in Figure \ref{fig:xray1}. At 10 cm from the tube exit, the maximum achievable dose rate is 4 Mrad/h, and the beam spot has a size of 8.7 cm, defined as the region where the dose rate is at least 80\% of its maximum value. The beam spot is significantly larger than the size of the ITkPixV1 chip. If the ITkPixV1 chip is placed at the centre of the beam spot, the delivered dose is expected to be uniform within 5\%.

One additional X-ray irradiation at a dose rate of 0.5 Mrad/h was performed at LBNL, using a Seifert X-ray system, but otherwise using the same setup for both X-ray machine settings and chip setup, and the dose rate was calibrated using the same calibration diode. 

In this paper, the TID received by the ring oscillators is quoted. This means the dose rate measured by the calibration diode is reduced by 20\% to account for the absorption of X-rays by metal layers covering the transistors in the chip. This correction factor was derived based on the X-ray spectrum and thickness of the metallisation layers and was previously studied in a dedicated measurement. Note that the results determined in this paper do not significantly depend on the assumptions made about the exact dose rate, as the main reported number are ratios of radiation damage at high and low dose rates. 

\subsection{Irradiation setup}

\begin{figure}
\centering
  \centering
  \includegraphics[width=0.45\textwidth]{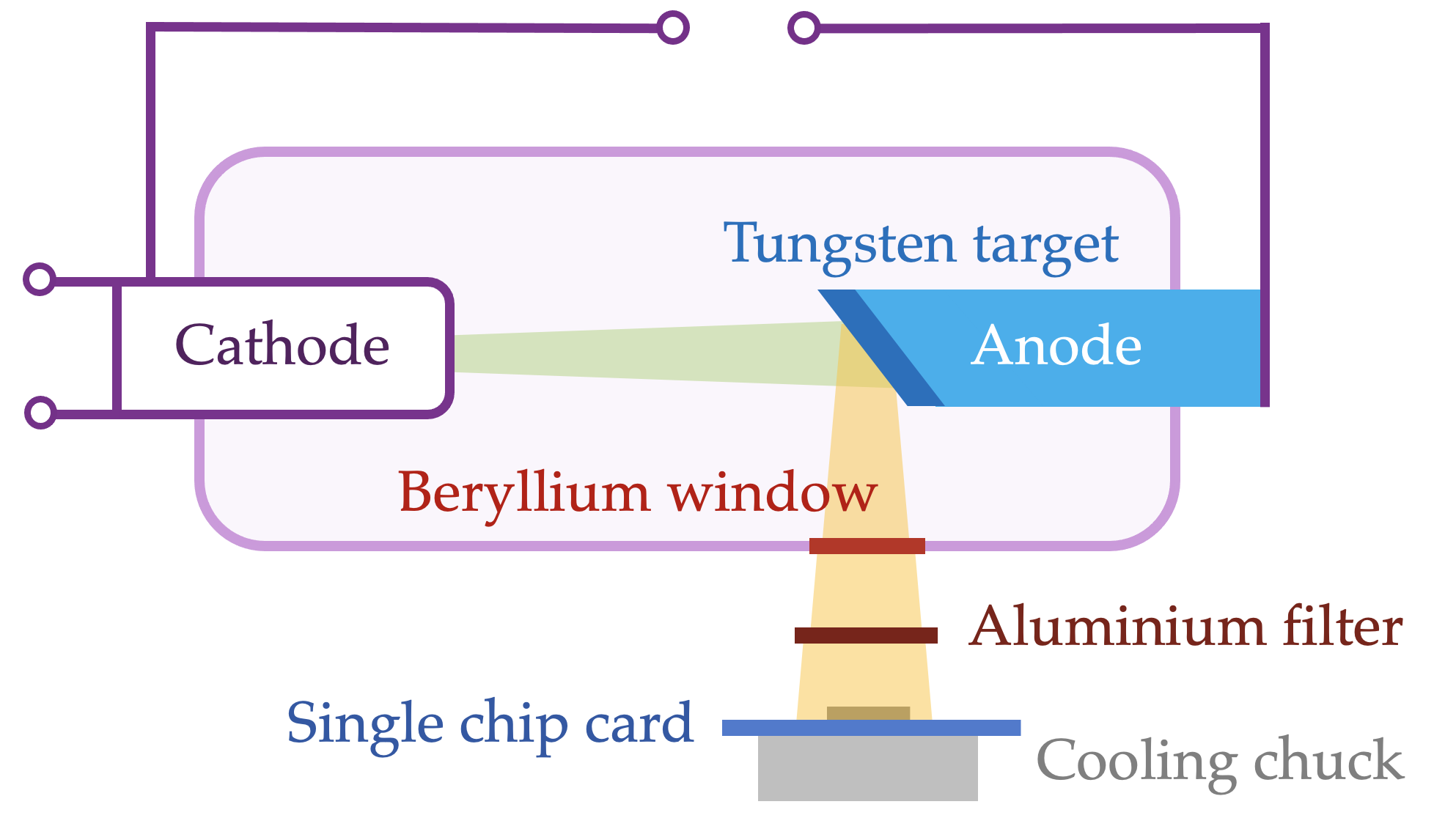}\qquad
  \includegraphics[width=0.45\textwidth]{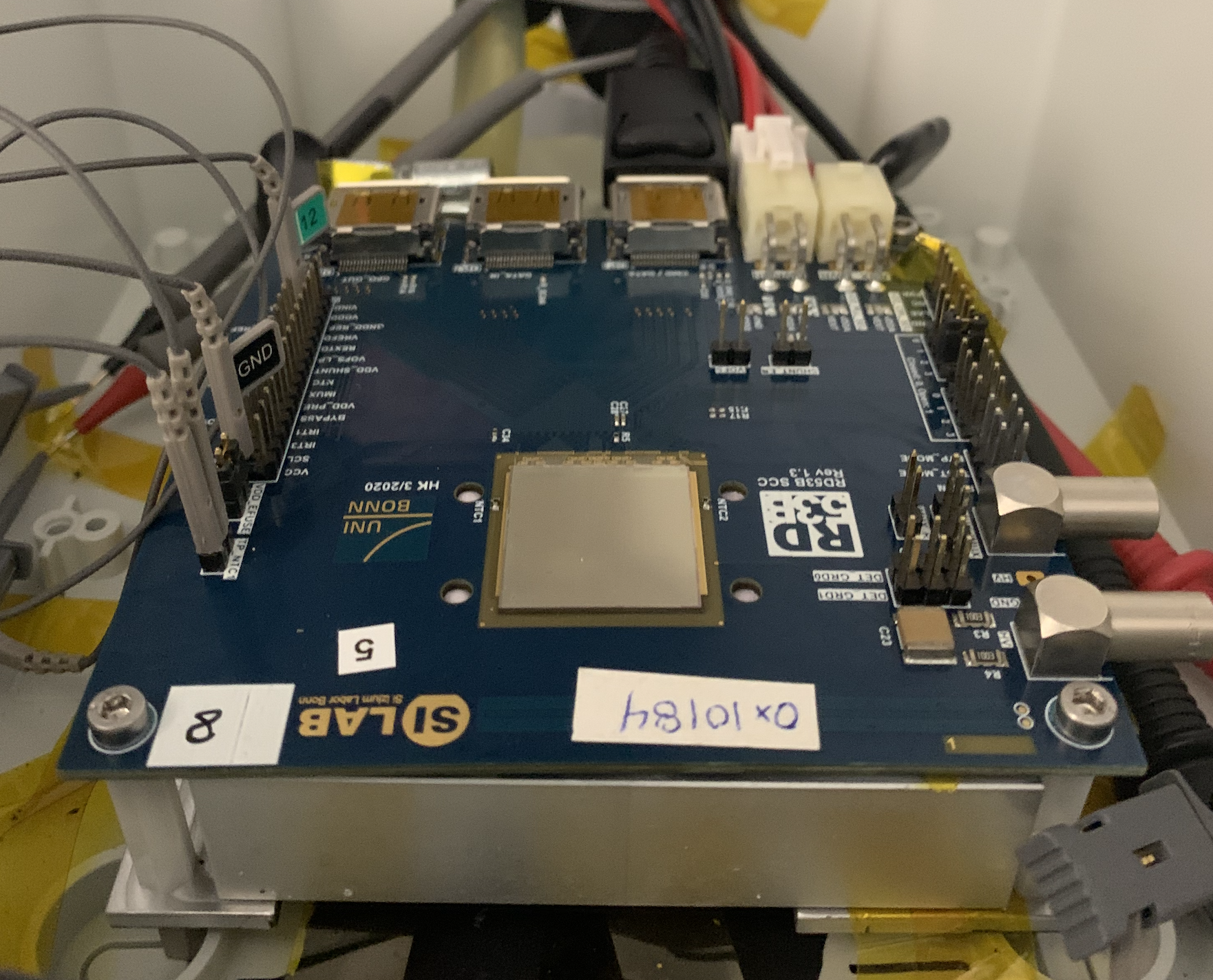}
  \caption{Schematic (left) and picture (right) of the setup used in irradiations of the ITkPixV1 chips. }
  \label{fig:setup}
\end{figure}

The ITkPixV1 chips were irradiated in the X-ray setup described above. The ITkPixV1 chip was mounted on a dedicated PCB (\textit{single-chip card}), which is used to supply the input voltage, connect to the YARR DAQ system \cite{YARR}, and has pins accessible for monitoring the voltages and currents in the chip. The X-ray tube was operated at a voltage of 40 kV, with a 150 $\mu$m Aluminium filter placed at the tube exit. The desired dose rate was achieved by adjusting the tube current and placing the chip at a certain distance from the tube exit. The setup is illustrated in Figure \ref{fig:setup}.

The irradiations were performed while the chip was cooled by placing the chip on a chuck connected to a Huber CC-508 chiller, which was able to keep the chip at a constant temperature of  -10$^{\circ}$C (+/- 1$^{\circ}$C), measured using an NTC thermistor on the chip PCB. This corresponds to a temperature of -7.5$^{\circ}$C on the chip, which is the maximum expected temperature during ITk operation. The maximum expected temperature was chosen, as radiation damage has been seen to be more severe at high temperatures. The temperature was monitored by reading out the resistance of an NTC thermistor on the single-chip card \cite{NTC}. The chip was powered by supplying 1.6 V to the chip. This voltage is distributed to the analog and digital domains of the chip via the chip internal linear regulators, leading to a nominal 1.2 V supplied on each. The voltage provided to the digital domain (\texttt{VDDD}) was monitored, as the frequency of the ring oscillators is sensitive to it.

The primary measurement during the irradiations was the frequency of the ring oscillators. The ring oscillators were read out at least after every 0.1 Mrad of delivered dose. When the ring oscillators were not measured, digital injection tests were run to simulate operation in a data-taking environment. Irradiations were performed at dose rates ranging between 20 krad/h and 30 Mrad/h. For the irradiation at a dose rate of 30 Mrad/h, the chip was placed as close as possible to the X-ray tube exit, and the beam spot was centred on the ring oscillators in the chip. For the other irradiation campaigns, the entire chip was irradiated.

\section{Results}

This sections outlines the calibration measurements performed on the ring oscillators, followed by the analysis and results of the X-ray irradiation campaigns. 

\subsection{Ring oscillator calibration measurements}

\begin{figure}
\centering
  \centering
  \includegraphics[width=0.45\textwidth]{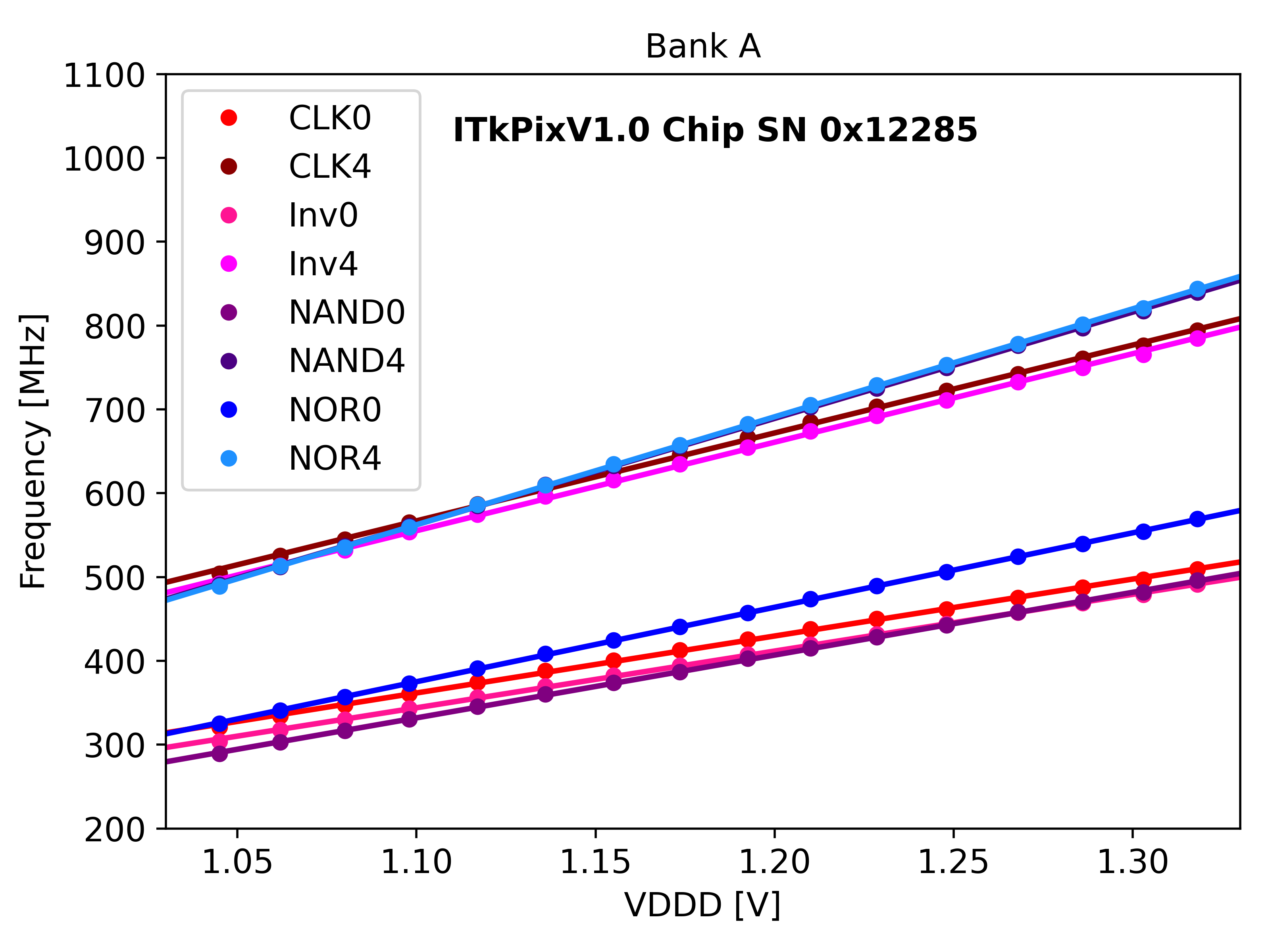}\qquad
  \includegraphics[width=0.45\textwidth]{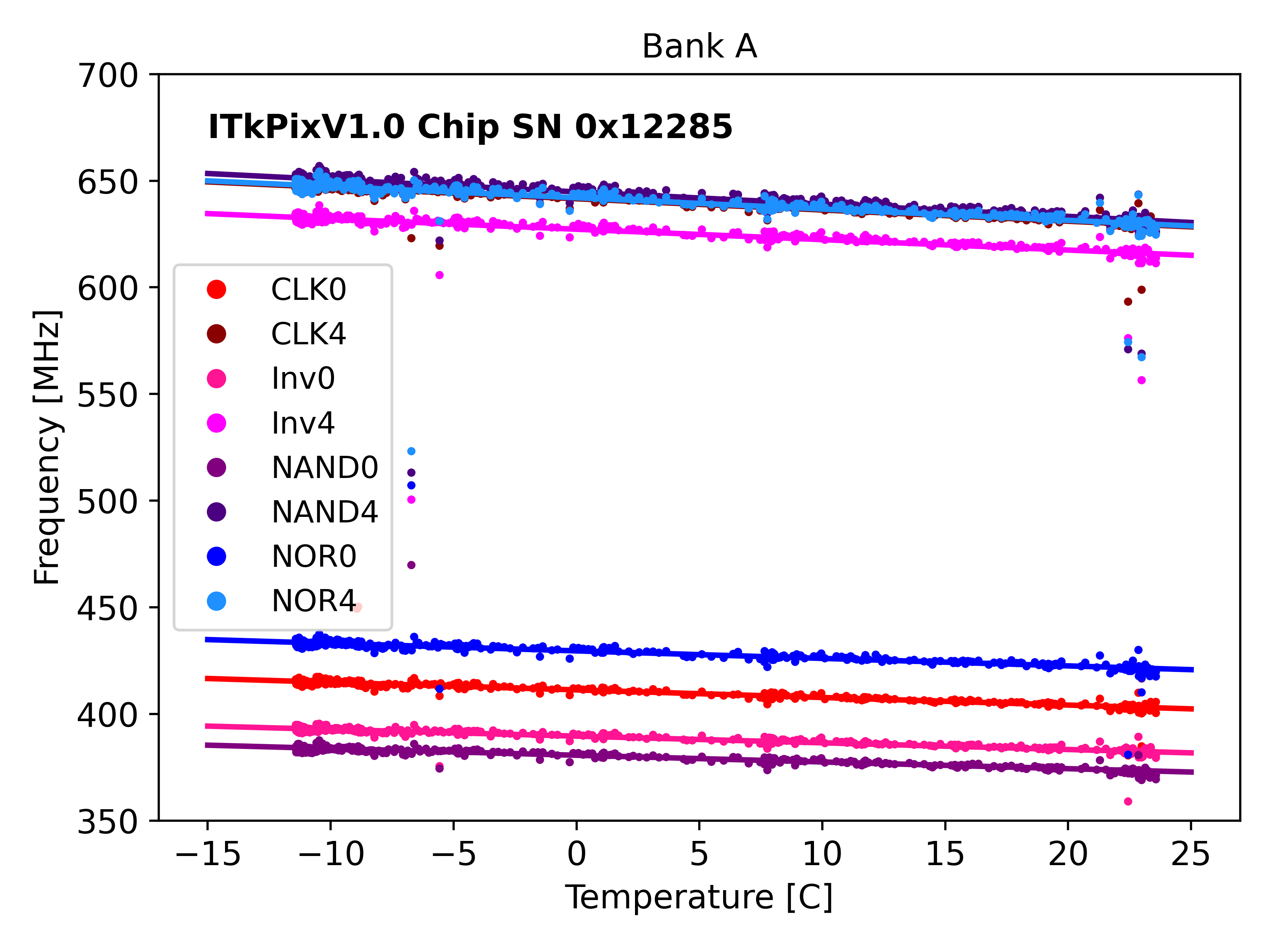}
  \caption{Examples of the VDDD (left) and temperature (right) calibrations of the ring oscillators in ITkPixV1 before irradiation. }
  \label{fig:ROSCcalib}
\end{figure}

Before performing irradiations, preliminary measurements of the ring oscillators were taken. There is a strong dependence of ring oscillator frequency on the supplied voltage and a slight dependence on the temperature. The dependence on the digital voltage was characterised for each chip before irradiation.

First, the frequency of the ring oscillators was measured while varying the \texttt{VDDD} voltage. The resulting dependence is shown in Figure \ref{fig:ROSCcalib} for a subset of the ring oscillators. A linear behaviour is expected and visible, and the ring oscillator frequency increases by up to 300 MHz when increasing \texttt{VDDD} from 1.0 to 1.3 V. The slopes and offsets are determined for each ring oscillator. They are then used to account for changes in \texttt{VDDD} during the irradiation.

Next, the dependence of the ring oscillators on temperature was measured. This was done by slowly cooling down the setup from room temperature to -12$^{\circ}$C and monitoring the ring oscillator frequency while accounting for any changes in \texttt{VDDD} using the parameterisations derived above. The results are also shown in Figure \ref{fig:ROSCcalib}. A clear linear frequency dependence on the temperature can be seen. The frequency changes by up to 20 MHz when changing the temperature from -12$^{\circ}$C to room temperature. The temperature dependence does not need to be accounted for during irradiation, as the temperature is kept constant.

\subsection{Dose rate dependence of TID damage}

\begin{figure}
\centering
  \centering
  \includegraphics[width=0.45\textwidth]{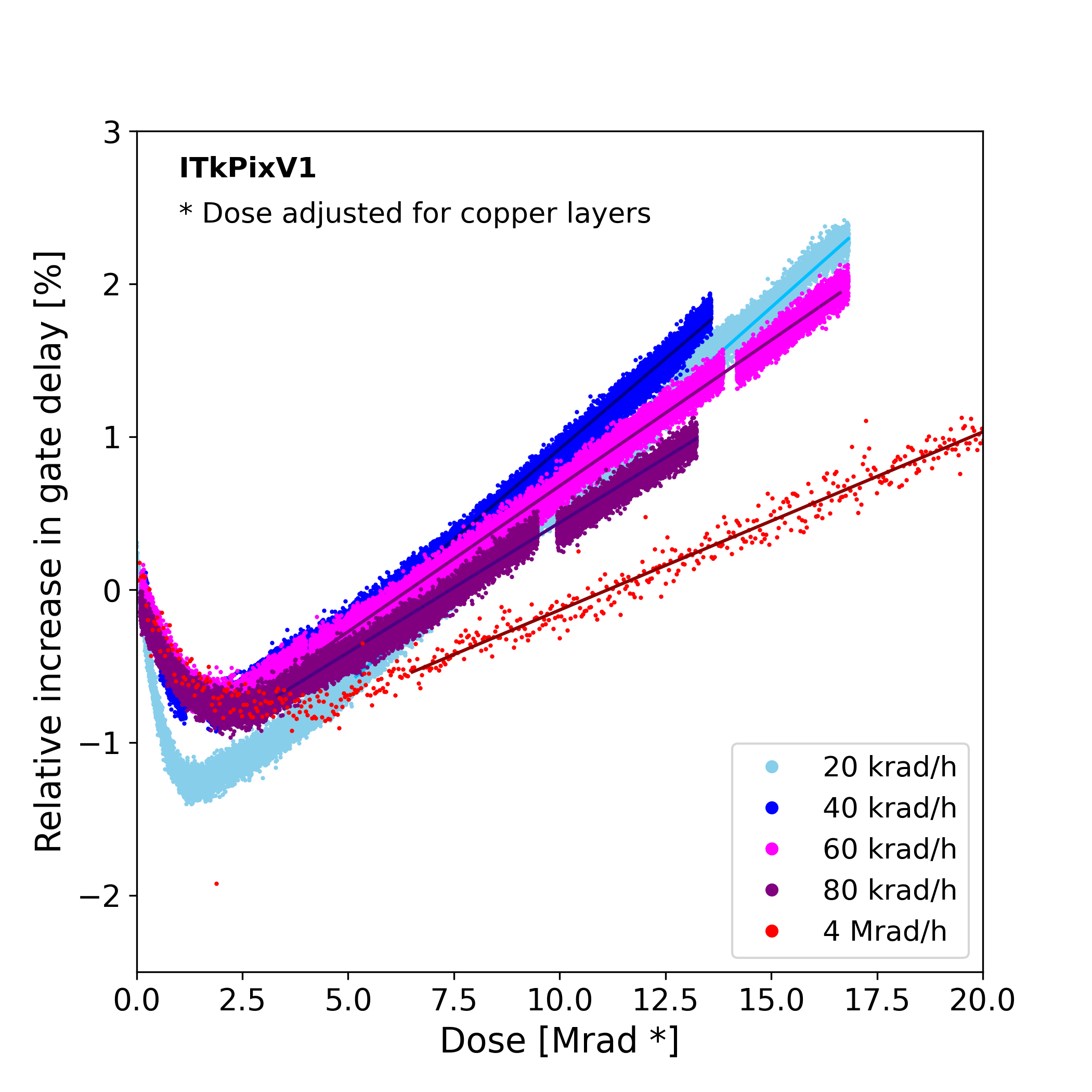}\qquad
  \includegraphics[width=0.45\textwidth]{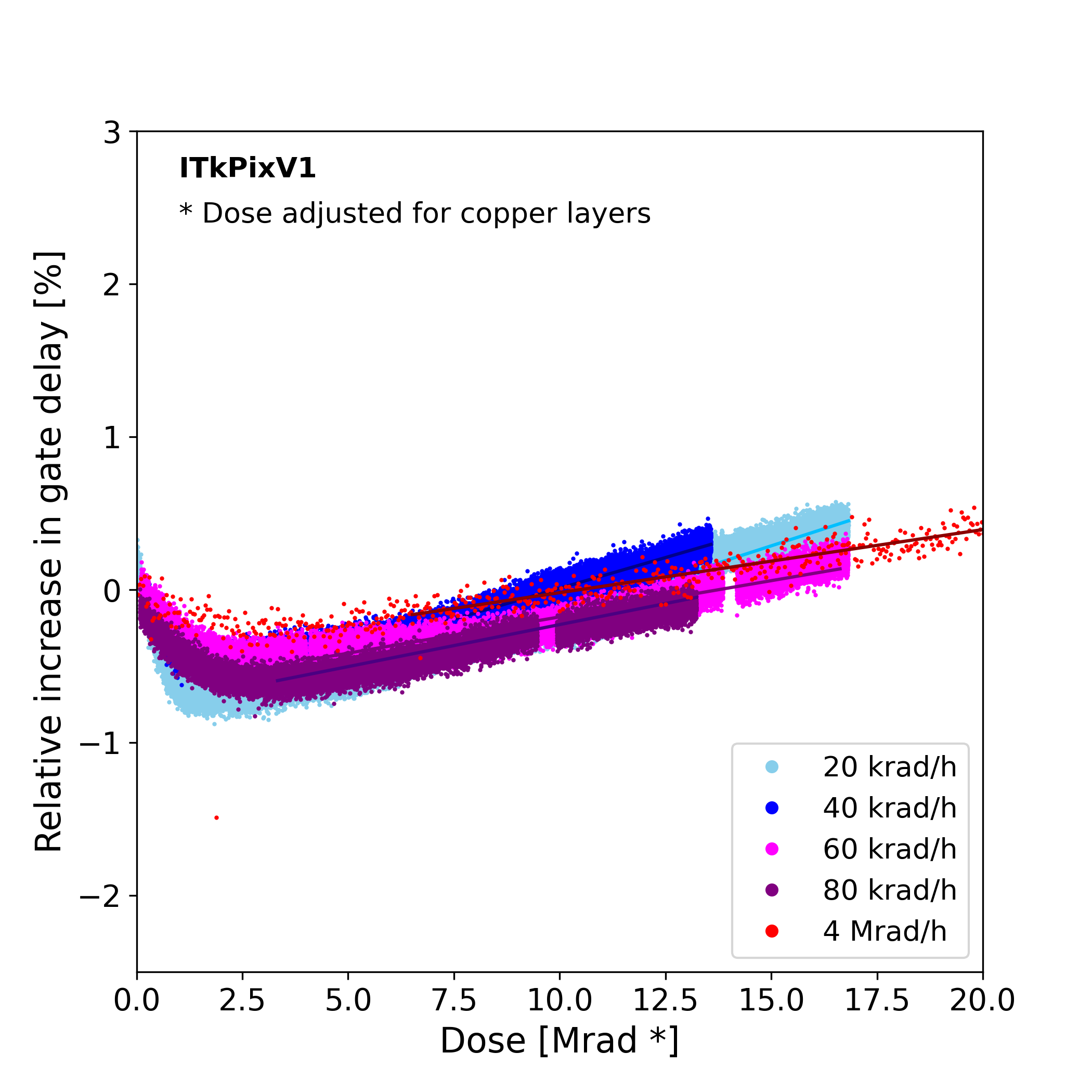}
  \caption{Gate delay increases at different dose rates for the ring oscillator made of Inverter gates, for the strength 0 (left) and strength 4 (right) gates.}
  \label{fig:RelativeDelay}
\end{figure}

Seven irradiation campaigns were performed, with dose rates of 20 krad/h, 40 krad/h, 60 krad/h, 80 krad/h, 0.4 Mrad/h, 4 Mrad/h and 30 Mrad/h, collecting at least 10 Mrad of TID during each irradiation campaign. The lowest dose rate of 20 krad/h corresponds roughly to the dose rate expected in the innermost layers of ITk. The frequencies of each ring oscillator were analysed and corrected for the variations in digital voltage (\texttt{VDDD}), using the calibration curves derived above. The ring oscillator frequencies were then translated into the delays of the individual gates, and the relative increase in gate delay was calculated. Figure \ref{fig:RelativeDelay} shows these relative increases in delay for the different irradiation campaigns for some examples of the strength 0 and 4 inverter gates. After an initial decrease in gate delay, corresponding to an initial drop of $V_{gs}$ in the transistor, the increase follows a linear trend, up to total doses of 20 Mrad. This trend is parametrised for each irradiation campaign by fitting a linear function to the data from a total dose of 3 Mrad onwards. The resulting fitted functions are also shown in Figure \ref{fig:RelativeDelay}.

\begin{figure}
\centering
  \centering
  \includegraphics[width=0.45\textwidth]{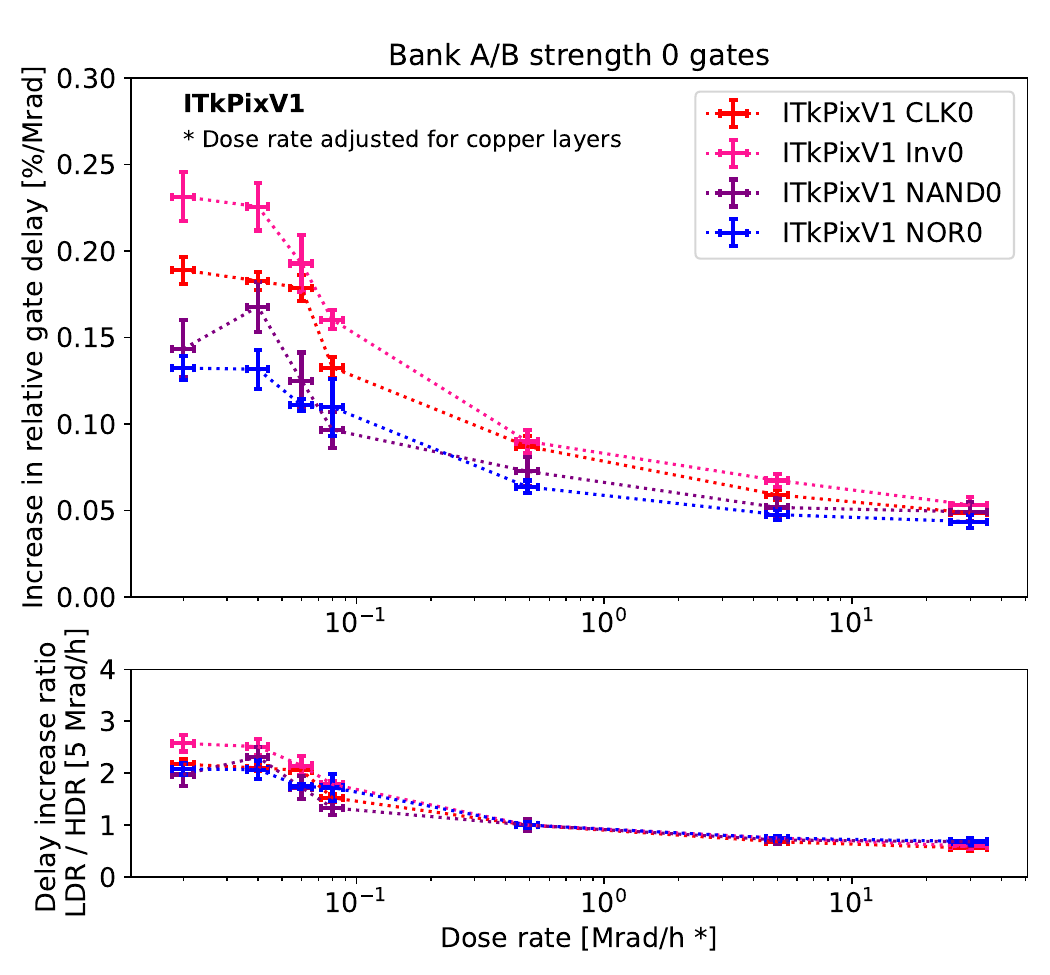}\qquad
  \includegraphics[width=0.45\textwidth]{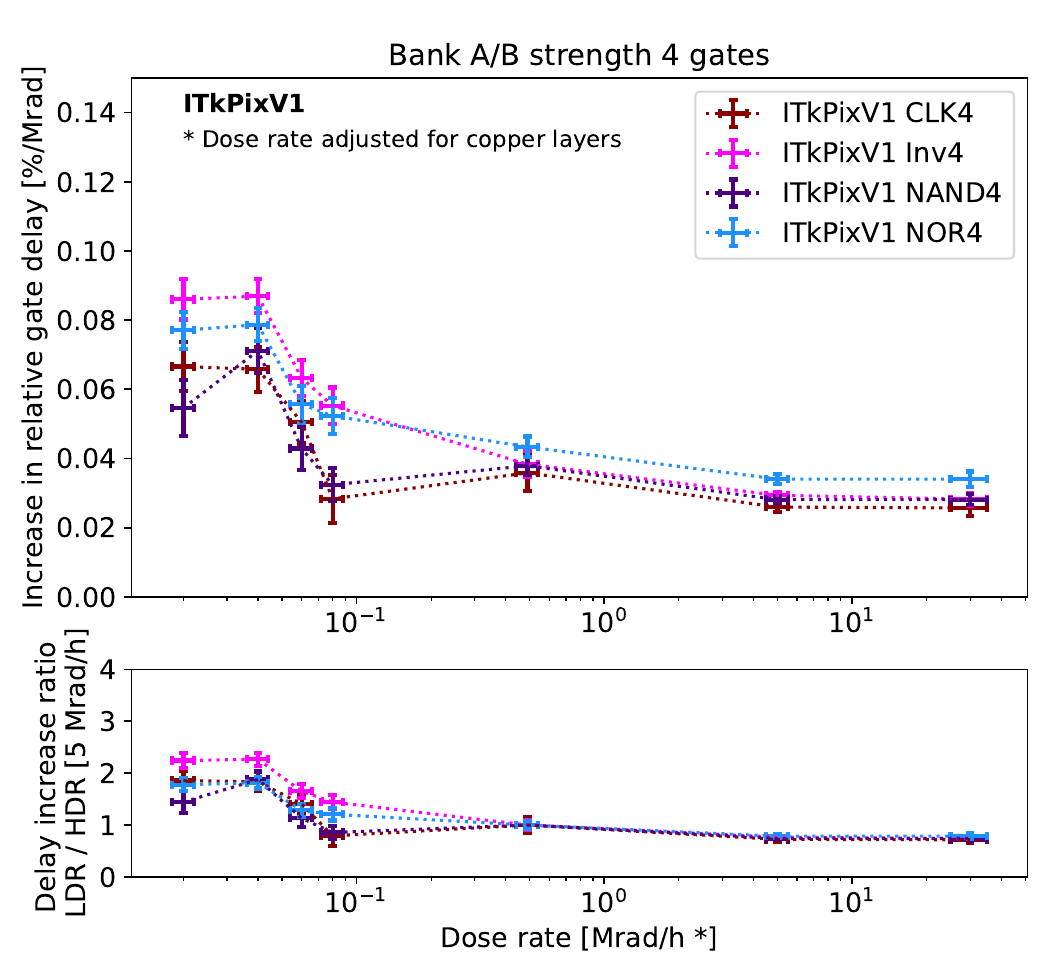}\\
  \includegraphics[width=0.45\textwidth]{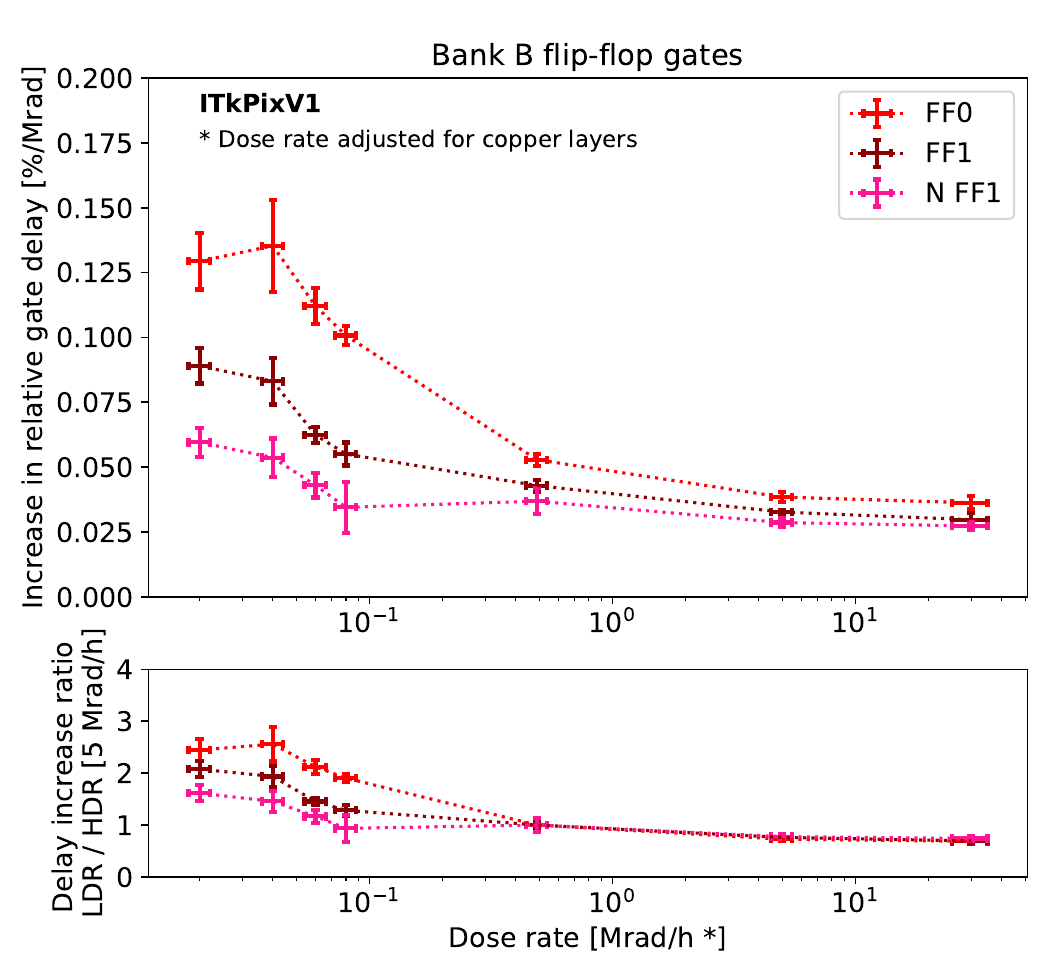}\qquad
  \includegraphics[width=0.45\textwidth]{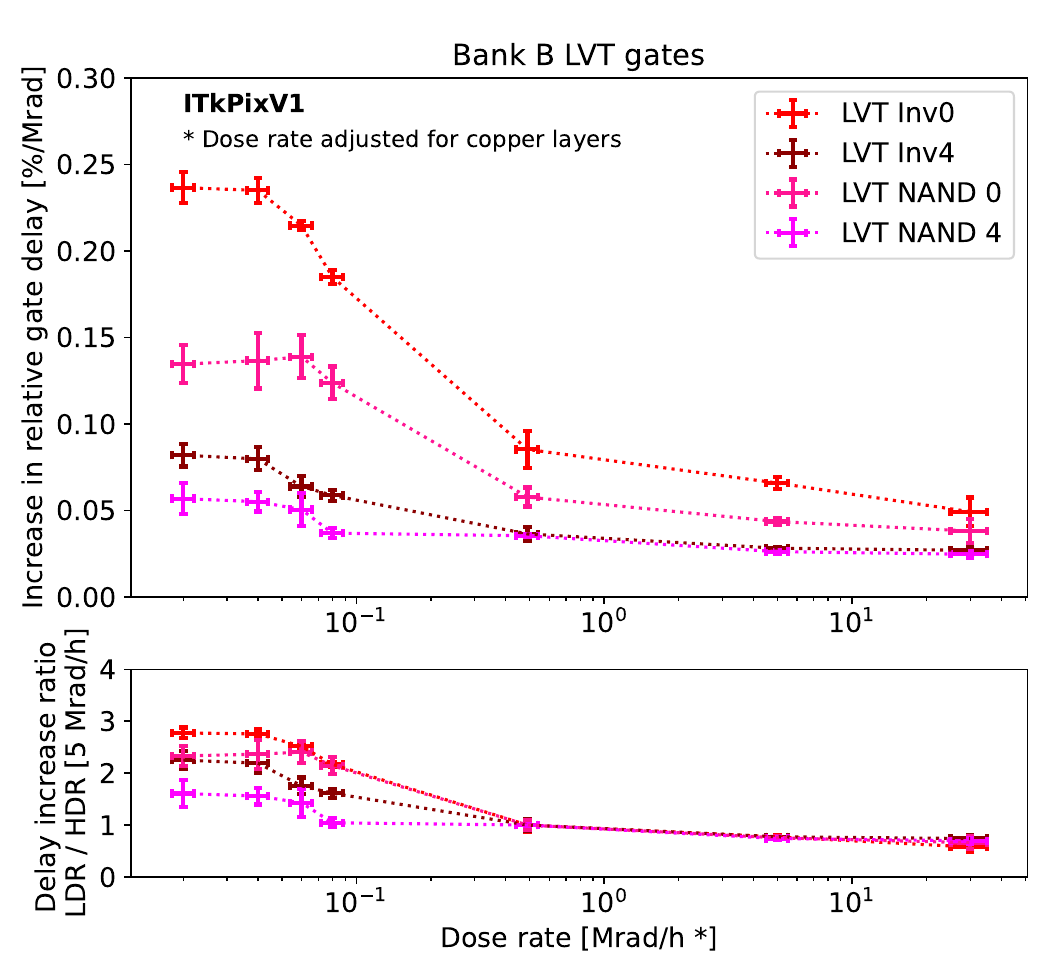}\\
  \includegraphics[width=0.45\textwidth]{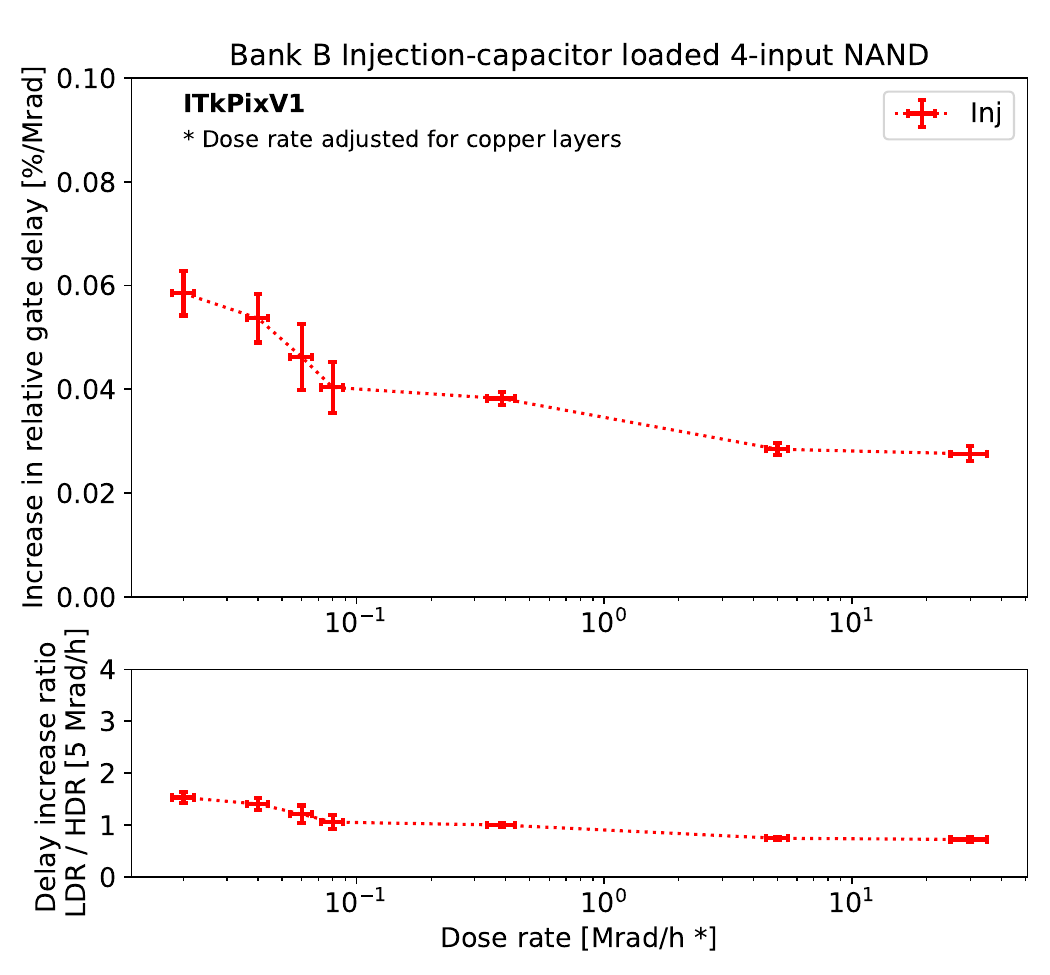}
  \caption{Fitted increase in relative gate delay per TID as a function of dose rate for all types of gates used in the ITkPixV1 ring oscillators. }
  \label{fig:Doserate}
\end{figure}

It is also possible to consider the increase in gate delay per TID as a function of dose rate. Figure \ref{fig:Doserate} shows the gate delay increase per TID (slopes of the fitted linear functions) as a function of the dose rate for each type of ring oscillator. The top panel in each plot shows the fitted values. The bottom panel shows the ratio of the values compared to the irradiation at a dose rate of 4 Mrad/h. This estimates the conversion factors needed for translating results from irradiations at high dose rates, commonly performed to high TID, and the expected gate delay degradation at the same TID delivered at a low dose rate. The results are averaged between the Bank A and Bank B ring oscillators of the same type. A systematic uncertainty on the dose rate is considered, depending on the precision with which it could be measured using the calibrated diode and the accuracy of the chip position in the beam spot. This uncertainty is usually of the order of 5-10\%. A systematic uncertainty on the fitted ring oscillator delays is also considered by varying the range used for the linear fit.

There is a clear trend of increased gate delay degradation at low dose rates. This increase depends on the transistor size, with the gate delay increase being roughly 3 times larger at low dose rates for the strength 0 gates, and only 2.5 times larger for the strength 4 gates. However, the conversion factor also depends on the exact type of gate used. It is also worth noting that the behaviour of increasing gate delay increase seems to flatten out at low dose rates, which is consistent with what was observed in single transistor measurements \cite{LDR}. This suggests that the gate delay degradation per TID will not be significantly worse at even lower dose rates, for instance, in the outer layers of ITk.


\section{Conclusions}

This paper characterised the dose rate dependence of radiation damage in 65 nm CMOS transistors in X-ray irradiation of the ATLAS ITk pixel pre-production readout chip ITkPixV1. Several X-ray irradiation campaigns were performed at dose rates varying between 20 krad/h and 30 Mrad/h. The resulting gate delay degradation was compared for total doses of at least 10 Mrad and more radiation damage was found for the same delivered TID at low dose rates, as expected from existing single transistor measurements \cite{LDR}. The exact increase in damage was found to depend on the size and type of gate used, and the gate delay degradation was observed to be 2-3 times worse at a dose rate of 20 krad/h, compared to 4 Mrad/h. 

This paper only covers dose rate effects at low total doses. To confirm the actual radiation hardness of the ITkPix digital logic, significantly longer irradiation campaigns at low dose rates are necessary, which require a different kind of irradiation method. Such an irradiation campaign with long exposure is possible using a Krypton-85 source and will be subject of a separate study. 

\section{Acknowledgements}

The research presented here was funded by a grant from the STFC (Oxford CG ST/S000933/1 and ST/W000628/1). Maria Mironova gratefully acknowledges the support of the STFC (grant number ST/S505638/1) and the Scatcherd Scholarship during her PhD. This work was supported by the Office of High Energy Physics of the U.S. Department of Energy under contract DE-AC02-05CH11231.


\bibliography{mybibfile}

\end{document}